\documentclass{article}
\usepackage{microtype}
\usepackage{graphicx}
\usepackage{subcaption}
\usepackage{booktabs} 
\usepackage{mdframed}
\usepackage{xcolor} 

\usepackage{hyperref}
\usepackage{enumitem}

\usepackage[preprint]{icml2026}

\usepackage{amsmath}
\usepackage{amssymb}
\usepackage{mathtools}
\usepackage{amsthm}
\usepackage{natbib}
\usepackage[dvipsnames]{xcolor}

\usepackage[capitalize,noabbrev]{cleveref}

\theoremstyle{plain}

\theoremstyle{definition}

\theoremstyle{remark}

\newcommand{\ignore}[1]{}

\usepackage[textsize=tiny]{todonotes}

\icmltitlerunning{We Should Separate Memorization from Copyright}

\begin{document}

\twocolumn[
  \icmltitle{We Should Separate Memorization from Copyright}



  \icmlsetsymbol{equal}{*}

\begin{icmlauthorlist}
\icmlauthor{Adi Haviv}{yyy}
\icmlauthor{Niva Elkin-Koren}{sch}
\icmlauthor{Uri Hacohen}{sch}
\icmlauthor{Roi Livni}{comp}
\icmlauthor{Shay Moran}{tec}
\end{icmlauthorlist}

\icmlaffiliation{sch}{Buchmann Faculty of Law, Tel Aviv University}
\icmlaffiliation{yyy}{Blavatnik School of Computer Science and AI , Tel Aviv University}
\icmlaffiliation{comp}{School of Electrical and Computer Engineering, Tel Aviv University}
\icmlaffiliation{tec}{Departments of Mathematics, Computer Science, and Data and Decision Sciences Technion– Israel Institute of Technology}
\icmlcorrespondingauthor{Adi Haviv}{adi.haviv@cs.tau.ac.il}

  \icmlkeywords{Machine Learning, ICML}

  \vskip 0.3in
]



\printAffiliationsAndNotice{}  

\begin{abstract}
The widespread use of foundation models has introduced a new risk factor of \emph{copyright issue}. This issue is leading to an active, lively and on-going debate amongst the data-science community as well as amongst legal scholars. Where claims and results across both sides are often interpreted in different ways and leading to different implications. Our position is that much of the technical literature relies on traditional \emph{reconstruction techniques} that are not designed for copyright analysis. As a result, memorization and copying have been conflated across both technical and legal communities and in multiple contexts. We argue that memorization, as commonly studied in data science, should not be equated with copying and should not be used as a proxy for copyright infringement. We distinguish technical signals that meaningfully indicate infringement risk from those that instead reflect lawful generalization or high-frequency content. Based on this analysis, we advocate for an output-level, risk-based evaluation process that aligns technical assessments with established copyright standards and provides a more principled foundation for research, auditing, and policy.
\end{abstract}

\section{Introduction}

Does the copying involved in developing and deploying GenAI constitute copyright infringement? Legal scholars disagree on this issue  \cite{bracha2023work,grimmelmann2015copyright,sag2020textmining} and recent court decisions have brought it to the forefront. Courts diverge on when and how copying during model training and deployment may violate copyright law \cite{GEMA_v_OpenAI_2024,Bartz_v_Anthropic2025,Kadrey_v_Meta_2025,Getty_v_StabilityAI_2025}.

Some courts have relied heavily on findings from computer science research (as in the \textit{GEMA} case) or on testimonies from technical experts (as in the \textit{Getty} case). Yet, there remains substantial conceptual ambiguity regarding the relationship between \emph{memorization} in generative models and \emph{copying} under copyright law. Many studies examine the extent to which models can reproduce training data - through extraction attacks, prefix completion, or strategic prompting \cite{nasr2023scalable,karamolegkou2023copyright,cooper2025extracting,cooper2025files} - without clearly distinguishing between \emph{memorization} as a technical property of a learning system and \emph{copying} as a legal notion tied to copyright infringement. As a result, the ability to reproduce training data is often implicitly treated as legally salient, even though the connection between these technical phenomena and the legal concept of copying is not settled.

Relatedly, a number of recent works explicitly investigate \emph{reconstruction attacks} that target copyrighted or potentially copyrighted content, with the aim of empirically characterizing the risks posed by generative AI models \cite{cooper2025extracting,yarkoni2025}. These studies provide valuable insight into the mechanisms and extent of memorization in modern models. At the same time, their findings underscore the need for conceptual clarity: the existence of a successful reconstruction attack, by itself, does not determine when, how, or whether memorization amounts to copying in the legal sense.

Altogether, both the legal literature and data science exhibit a keen interest in further understanding and mitigating this effect. However, the two disciplines do not always align: they differ in their assumptions, objectives, and standards of evidence, resulting in largely parallel yet weakly integrated bodies of work. Our focus in this position is on the gap we observe between how reconstruction attacks are commonly interpreted and how copying is interpreted across the two disciplines. Overall, we aim to provide further clarification on the following question:

\begin{center}
To what extent can \emph{memorization} be equated with \emph{copying}?
Which forms of memorization are legally relevant, and what can they demonstrate in copyright analysis?
\end{center}

\textbf{This paper takes the position that memorization, as it is traditionally defined and analyzed in machine learning, is relevant to the notion of copying but is neither equivalent to it nor sufficient or necessary for establishing copyright infringement. We argue that recent technical and legal literature frequently conflates these notions, and that disentangling them is essential for developing principled approaches to mitigating copyright-related concerns in generative models.}

Towards this goal, we advance two central claims regarding the interpretation of memorization and reconstruction attacks in the context of copyright law.

\textbf{First}, we argue that the design and interpretation of reconstruction and extraction attacks must carefully distinguish between \emph{proof of memorization} as a technical property of learning systems and \emph{proof of copying} as a legal concept under copyright law. While many existing evaluations implicitly treat memorization of copyrighted material, as legally meaningful, this assumption obscures the fact that copyright law attaches significance to specific forms of reproduction under particular conditions. Properly assessing copyright risk therefore requires a clearer understanding of what constitutes copying in the legal sense, and how - if at all - technical notions of memorization map onto it.

\textbf{Second}, we challenge a prevalent viewpoint that conflates memorization and copying \emph{within the model}. We argue that the successful derivation of copyrighted content from a generative model does not, by itself, establish that the model holds a copy of the data. Such inference overlook the central role (volition act) of the user in the generation process: its initiation, the specificity of the prompt used, and the amount of prior information supplied. While it is uncontroversial from an information-theoretic perspective that a trained model encodes information about its training data, this does not imply that the data is stored or represented in the model in a database-like manner, nor that the model itself constitutes a copy of that data in the legal sense.

To support these claims, we make the following contributions. We first review the elements of copyright law most relevant to generative models, clarifying what is protected and what constitutes infringing reproduction. We then develop a taxonomy that organizes the principal points of friction between technical notions of memorization and copyright doctrine, identifying two central loci of dispute: claims concerning the legality of the model itself, and claims concerning the presence of copyrighted material in model outputs. Finally, we reinterpret existing memorization and extraction studies through these legal lenses, identifying which technical signals plausibly indicate copyright risk and which do not, and characterizing the forms of generated outputs that may implicate copyright across modalities, distinguishing literal from non-literal reproduction and thin from thick protection.

Our goal is to steer the research community toward a more principled and legally grounded approach.  We advocate for evaluation frameworks that focus directly on copyright-relevant harms at the output level, align technical metrics with copyright doctrine, and support  mitigation strategies that target genuine infringement risks rather than abstract concerns about memorization. By clarifying what copyright law actually protects and when infringement occurs, we aim to support more effective and legally informed practices in the development and deployment of generative models.

\section{Copyright Background}
\label{sec:copyright_background}

\begin{figure}[t]
    \centering
    \includegraphics[width=\linewidth,trim=0 0 2 0,clip]{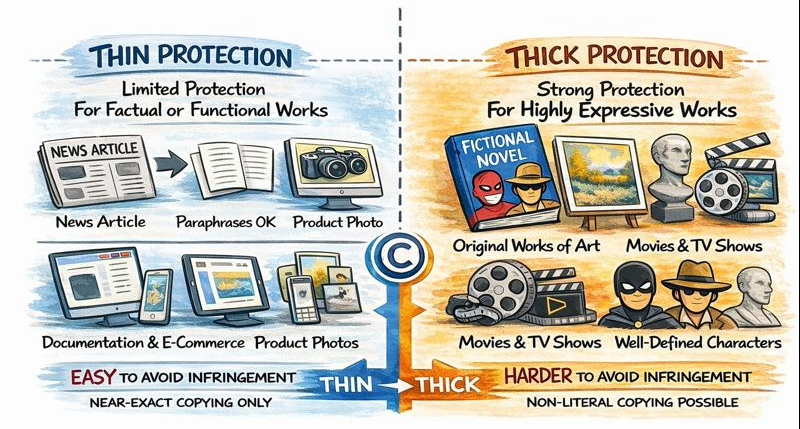}  
    \caption{Illustration of the spectrum of copyright protection, ranging from thin to thick.}
     \vspace{-0.3cm}
    \label{fig:thin_v_thick_vis}
\end{figure}

We begin with a brief and non-exhaustive overview of copyright law, focusing on its purpose, scope, and the legal elements required to establish infringement. This legal framework serves as a foundation for identifying when copying during  the development or deployment of generative models may constitute  unlawful  reproduction under copyright law.

\subsection{Copyright Purpose and Principles}

Under U.S. law, copyright seeks to incentivize the creation of expressive works for the benefit of society by granting authors a limited set of exclusive rights in their original works   \cite{us_constitution,mazer1954}.  At the same time, because creative works function as inputs for future creation, learning, and transformation, overly expansive protection risks undermining copyright’s very purpose. Copyright therefore operates through a calibrated structure of partial exclusivity and public access, granting authors control over certain uses of their works while leaving other aspects and uses unencumbered, thereby balancing private reward with collective creative progress \cite{litman1990}.  \emph{Crucially, facilitating public access to data is not in conflict with copyright law but instead reflects one of its core organizing principles.} 

In the context of generative AI, these foundational objectives are manifested in the tension between the rights of copyright holders in their original expressions and the permissible scope of data use during model training and deployment.

\subsection{ Which Copying Constitutes a Copyright Infringement? }
\label{subsec:whichcopyingconstitutes}

 Under U.S. copyright law, copyright \textit{infringement} occurs when a person violates any of the exclusive rights granted to a copyright owner \cite{usc106} unless the use falls within exceptions and limitations provided by law (most notably fair use) \cite{usc107}.  In the context of GenAI, copying may constitute copyright infringement if the following three conditions are satisfied:
 
 \begin{enumerate}[leftmargin=10pt, nosep]
\item \textbf{Protected Work}: The material copied must be an original expression protected by copyright law, as opposed to unprotected elements (e.g., facts, ideas, public domain content as elaborated in ~\cref{subsec:whatis} and Appendix~\ref{apx:copyright_Infringement_background}).

\item \textbf{Copying}: Copyright infringement requires reproducing a work, namely fixing it in a material object from which it can be "perceived, reproduced, or otherwise communicated" including with the aid of a machine. Liability further depends on \textit{access} to the original work and \textit{substantial similarity} to its protected elements, and subsection~\ref{subsec:whatis} examines how the scope of protection varies with the amount of protected expression present. 

 \item \textbf{No applicable Exceptions or Limitations}: The use does not qualify for any legal exceptions such as temporary copying, educational use, other statutory exemptions, or \textit{fair use} (i.e., a privilege that permits certain uses under a flexible, case-by-case standard. Described in App.~\ref{apx: fair_use}).  
\end{enumerate}
Further details are provided in Appendix ~\ref{apx:whichcopyiscopyright}.

\subsection{What is protected under copyright, and what is not}
\label{subsec:whatis}

Copyright operates across all fields of creative endeavor, but within each domain it protects only certain elements - those that qualify as original expression - while explicitly excluding others, as provided by the U.S. Copyright Act. 

Per the U.S.\ Copyright Act \cite{usc102b}:
\begin{center}
\emph{``In no case does copyright protection for an original work of authorship extend to any idea, procedure, process, system, method of operation, concept, principle, or discovery, regardless of the form in which it is described, explained, illustrated, or embodied in such work.''}
\end{center}

Copyrighted works often combine protected expression with unprotected elements. To determine whether protected expression has been copied, courts apply a filtration analysis that removes unprotectable material and evaluates similarity only with respect to the remaining original expression \cite{ComputerAssoc_v_Altai_1992}. We expand on unprotected subject matter in Appendix~\ref{apx:copyright_Infringement_background}.

Filtration is conducted at the time of the alleged infringement and is dynamic. Elements that were once expressive may lose protection if they become widely adopted and function as industry standards,  interoperability requirements, or cultural genres (scènes à faire). \cite{hacohen2024,menell1989}. For instance, widely adopted interface declarations that serve functional rather than expressive purposes may lose protection \cite{googleoracle2021}. Conversely, some expressive constructs, most notably fictional characters, may gain protection over time as they become more fully developed \cite{dccomics2015towle}. After filtration, the amount of protected expression that remains determines the thickness of copyright protection (Figure~\ref{fig:thin_v_thick_vis} illustrates the thin–thick protection spectrum.):

\begin{itemize}[leftmargin=10pt, nosep]
 \item \textbf{Thin copyright protection.} Works containing only a limited range of protected expression receive thin protection, which extends solely to literal or near-identical copying and does not reach remote non-literal similarity. Thin protection typically applies to works dominated by factual, functional, or conventional elements. News articles that primarily convey facts, for example, are protected only against verbatim copying, not against alternative descriptions of the same information \cite{feist1991, applemicrosoft1994}. Similarly, standard product or profile photographs used for commercial purposes generally receive only thin copyright protection, extending to exact duplication but not to independently created photographs of the same subject \cite{EtsHokin_v_SkyySpirits_2003}.
 
 \item \textbf{Thick copyright protection.} Works embodying a substantial concentration of expressive choices receive thick protection, which extends beyond literal copying to non-literal appropriation of protected expressive elements. Fictional novels with well-developed characters and narrative structures fall squarely within this category. Even significant rewording may infringe if it reproduces distinctive characters, relationships, or narrative architecture \cite{applemicrosoft1994}. The same principle applies to sufficiently delineated fictional characters, whose protection extends across stories and formats \cite{dccomics2015towle}. We elaborate more on copyright scope and infringement in Appendix~\ref{apx:copyright_Infringement_background}. 
 Finally, regardless of thickness, uses of protected works may be non-infringing if they fall within statutory limitations or exceptions, most notably fair use \cite{usc107}, which permits certain socially valuable uses such as criticism, research, and parody \cite{campbell1994acuffrose}.

\end{itemize}

\section{Memorization and Copying}

 Developing and deploying generative models involves multiple types of copying in the physical sense, which raise different types of copyright infringement claims in the legal sense: copying in the course of training, copying in the model itself, and copying in the generated output. 
 
\subsection{
 Copying in the course of training}

The point of contention that is arguably least relevant to memorization concerns copying undertaken in the course of training. This category includes the reproduction of copyrighted works into digital files and the creation of training datasets for the purposes of curating, cleaning, and preparing materials for model training. The concern here largely mirrors traditional questions surrounding the digitization of copyrighted works. For the sake of completeness, and because this issue is sometimes conflated with other points of contention, particularly our second point, we address it briefly here. 

Courts and commentators generally agree that traditional digitization constitutes legal copying, \cite{UMG_v_MP3com_2000}, but that such copying is likely permissible under fair use when undertaken to train generative models \cite{usc107}. 

In \citet{Bartz_v_Anthropic2025}, for example, the court held that copying books to train Claude constituted fair use, as did the digitization of lawfully purchased print books to create space-saving and searchable digital files, where Anthropic did not add new copies, create derivative works, or redistribute the originals \cite{Kadrey_v_Meta_2025}. Many copyright scholars also endorse the view that copying for the purpose of model training constitutes fair use \cite{lemleycasey2021, samuelson2024fairuse, sag2023fairness, gillotte2019copyright, lim2018ai, grimmelmann2015copyright}. 

\subsection{Is The Model a Copy?}\label{subsec:themodel}

The second point of contention invokes the concept of memorization to argue that generative models are themselves a copy of the data used for training them because that data is encoded in the model's weights, from which it can be "perceived, reproduced, or otherwise communicated." \cite{cooper2025files, usc101}. A recent decision by a German court adopted this view \cite{GEMA_v_OpenAI_2024}. 


\citet{cooper2025files} endorse this view, relying on memorization-implying data regurgitation attacks. They claim that, "a model has 'memorized' a piece of training data when (1) it is possible to reconstruct from the model (2) a (near-)exact copy of (3) a substantial portion of (4) that specific piece of training data.". They argue that this definition may satisfy the threshold for a \textit{copy} under U.S. and E.U. copyright law \citet{Dornis_Lucchi_2025_GenAI_EU_Copyright}. On the other hand, \citet{bracha2023work,bracha2025} challenges this view for focusing on the \textbf{physical dimension} of copying, which, he argues, falls outside the expressive domain protected by copyright.  Similarly, others contend that merely technical reproduction is non-communicative, and therefore non-use of the work “as a work” \cite{Drassinower_2015_Whats_Wrong_with_Copying, Strowel_2018_Reconstructing_Reproduction}.
Building on that, many legal scholars argue that copyright policy should shift away from the technical analysis of copying, explicitly excluding “non-expressive” or “non-consumptive” uses, such as those typical in machine learning or text mining \cite{sag2020textmining,Craig_2025_AI_Copyright_Trap,Litman_2017_Fetishizing_Copies}.

Our focus here is on the technical claims, and as we explore in the next two paragraphs, we claim that existing data reconstruction attacks do not necessarily inform us if a ``copy" of the data exists in the model.

\paragraph{What is provably known about information in the model}

The information between model and sample is often studied in the context of generalization theory  \cite{tishby2000information, mcallester1999pac, xu2017information,bassily2018learners, floyd1995sample} as well as very strict models of privacy \cite{kasiviswanathan2011can}. 
Such information is often considered undesirable. It is a precursor for \emph{overfitting} \cite{xu2017information, mcallester1999pac, bassily2018learners}, and serves as a threat to privacy \cite{kasiviswanathan2011can}. As such, irrespective of copyright issue, its mitigation has been studied. 

Perhaps counterintuitively, works have shown that such memorization may be \emph{necessary} for learning
\cite{livni2023information,attias2024information,feldman2020does, feldman2025trade, brown2021memorization}. This line of work demonstrate that, even in simplistic problems, a form of mutual information between model and data is \emph{necessary} to solve learning tasks.
But importantly, there are two issues with such theoretical results that are relevant to the question of copyright. 

First, some results \cite{feldman2020does, attias2024information, livni2023information} demonstrate that \emph{some} information exists, but can't tell us \emph{what} information exists. As such, since the mathematical formalism can't make a distinction between \emph{expression} and \emph{non-expressive elements} as required for copyright analysis, their implications are limited.

The second issue, which is further discussed in the next paragraph, is that these results are \emph{conditional}. For example, \citet{brown2021memorization} shows an instance where almost \emph{all} the information exists in the model. However, this is conditioned on the distribution, or prior, of the data. In a nutshell, it means that information can be extracted but only by an attacker that has a prior on the distribution of the data.

\paragraph{Are existing attacks demonstrating copying}
Attacks in practice are always conditioned on auxillary data, and at least one court acknowledged that as a limitation \cite{Getty_v_StabilityAI_2025}.

Prior, when courts applied the definitions for copyright to earlier digital technologies, they focused less on the mere presence of information and more on its retrievability. A PDF file or MP3 which are considered a copy of the original work can be reproduced exactly using standard software. They are not a copy merely because they hold the information over the data. For example, an encrypted PDF file to which no one holds the secret key arguably does not constitute a copy, even though it has exactly the same amount of information on the file as the non-encrypted file. 

Crucially, model’s internal weights are not directly interpretable, and memorization is inferred from attacks rather than directly observed. 
Attacks are procedures where we intentionally interact with the machine in order to demonstrate that some information exists in it. They are designed for privacy and security purposes. This observation has been noted by \citet{carlini2025privacy} that discusses three important distinctions: such works focus on \emph{personal information}, they don't reflect \emph{average case behavior} and they report lower bounds and not estimates on information. 

But importantly, even when the attacks are applied to copyrighted works, such attacks do not necessarily demonstrate reproduction. Distinctive from PDF reader, reconstruction attacks are not standard softwares that ``decode" information from weights. They are noisy, unreliable procedure that incorporates prior knowledge and intent.
Consequently, reconstruction attacks and information-theoretic analyses do not,  for example, distinguish between a hard drive that stores an explicit copy of a file and one from which the file has been deleted but can still be reconstructed using prior knowledge, invasive techniques, and active effort.

In our context, for example, \citet{cooper2025extracting} extract passages from the Great Gatsby by providing prompts from the Great Gatsby. A reconstruction attack by \citet{carlini2021extract} uses access to the training data, and targets specific, anomalistic prompts, intentionally.  Similarly, the plaintiffs in the recent case of \citet{Getty_v_StabilityAI_2025}, were able to extract the Getty Images' trademark by replicating the exact grammar and structure of real Getty Images' captions.

The issue is further illustrated by studies reconstructing images from brain activity \cite{beliy2019voxels, wen2018neural, shen2019deep, shen2019end}. They show that fMRI signals contain information about observed images. However, demonstrating the presence of information does not establish that a “copy” of the image exists in the brain. Such reconstructions rely on highly non-neutral attackers, pretrained on natural images and actively filling in missing or ambiguous signal components that cannot be directly interpreted.

The availability of information which enables copying needs to be distinguished from the actual act of preparing  a copy in the copyright legal sense. Courts have reiterated this principle, holding that “to establish direct liability … something more must be shown than mere ownership of a machine used by others to make illegal copies” \cite{CoStar_v_LoopNet_2004}.  In the context of \textit{Cablevision}, the court held that subscribers, not the service provider, fixed the copies by selecting programs to be recorded. Reproduction, the court reasoned, requires volitional conduct: the user “presses the button,” while the system merely “automatically obeys commands.”

Generative AI models are not recording machines. But nevertheless, existing attacks are limited in what they show. They demonstrate that a model holds data from which an infringing copy might be extracted with volitional additional effort. Copyright case law has consistently cautioned against imposing direct liability for copies made incidentally through automated processes absent volitional conduct. In \citet{RTC_v_Netcom_1995}, the court held that even automated, temporary copies made by an Internet service provider do not constitute direct infringement without a volitional act, explaining that “the mere fact that Netcom’s system incidentally makes temporary copies of plaintiffs’ works does not mean Netcom has caused the copying.” This reasoning became foundational for later protections afforded to online intermediaries. Accordingly, reconstruction attacks illuminate only a limited aspect of the overall infringement analysis.

\subsection{Copying in the generated output}\label{subsec:theoutput}

Finally, outputs generated by a model that are substantially similar to works in its training dataset may themselves constitute legal copies, whether reproductions or derivative works. Courts and commentators largely agree with this premise, although no court has yet ruled on it decisively, largely because plaintiffs have so far failed to produce sufficient evidence of actual infringing outputs \cite{Bartz_v_Anthropic2025,Kadrey_v_Meta_2025}.

Nevertheless, the copying of generated outputs remains a genuine concern, and further work aimed at identifying and characterizing the associated risks is an important direction for future research. However, as we elaborate in the next section, our position is that existing attacks should be aligned with the relevant legal definitions, and that their effectiveness should be assessed with respect to these criteria.
\section{What Constitutes Copyright Infringement in Generative Outputs}
As discussed in Section~\ref{sec:copyright_background}, copyright liability turns on model outputs and requires access, substantial similarity of protected expression, and the absence of defenses such as fair use. The following subsections map existing extraction and reproduction attacks onto these criteria. We emphasize that many of these works explicitly refrain from legal claims and that our concern lies not with the attacks themselves, but with how their results are often interpreted downstream by courts, policy actors, and secondary literature.

\subsection{Thin Protection: Factual or Weakly Expressive}

When the underlying data is dominated by factual or functional elements, the scope of protection is narrow, and the threshold for similarity to a training example for infringement is high. Liability is strictly bound to cases of verbatim or near-verbatim duplication.

\paragraph{Verbatim attacks} Near-duplicate extraction attacks directly test this failure mode. In diffusion models, prior work shows that carefully designed prompts can elicit images that closely match specific copyrighted photographs from the training data, replicating pose, lighting, background, and framing, i.e., the photographer’s protected expressive choices \citep{somepalli2022diffusion,carlini2023diffusion}. These attacks are verified by retrieval against training sets using perceptual similarity over DINO or CLIP embeddings, with a high threshold (i.e., the images are perceptually indistinguishable to human observers). This reflects the goal of detecting the same expressive instance under minor transformations rather than syntactic similarity.

A parallel line of work exists for language models. \citet{carlini2021extract} show that prefix-based prompts can trigger long verbatim continuations, often exceeding 50 to over 100 consecutive tokens, detected using longest common subsequence and n-gram overlap (The average words per token are ~0.75 words, meaning 100 tokens are $\approx$ 75 words). \citet{nasr2023scalable} scale this approach by indexing billions of n-grams and recovering long contiguous copyrighted passages from production LLMs, frequently spanning hundreds or thousands of tokens.

Across these works, similarity is predicated on exact token sequences rather than abstract or semantic resemblance. These attacks, therefore, test literal reproduction of protected expression rather than reuse of ideas or narrative structure. For works dominated by unprotected elements, this is precisely the condition required for infringement (conversely, works dominated by protected elements enjoy even broader protection, far beyond such literal copying). 

\paragraph{Memorization does not always indicate copyright risk} Importantly, not all memorization-style attacks correspond to copyright-relevant copying. Several probing methods primarily reveal training-set membership or factual recall rather than protected expression. \citet{chang2023speak} elicit book titles, author names, character names, and short quoted fragments, but find that long verbatim passages are rare. \citet{zhao2024pip} reconstruct only short spans using partial-information prompts, and earlier work extracts uncopyrightable factual strings such as phone numbers or keys \citep{carlini2019secret}. These results demonstrate exposure and memorization, but not infringement under a verbatim-reproduction standard.
\begin{figure}[t]
    \centering
\includegraphics[width=0.97\linewidth]{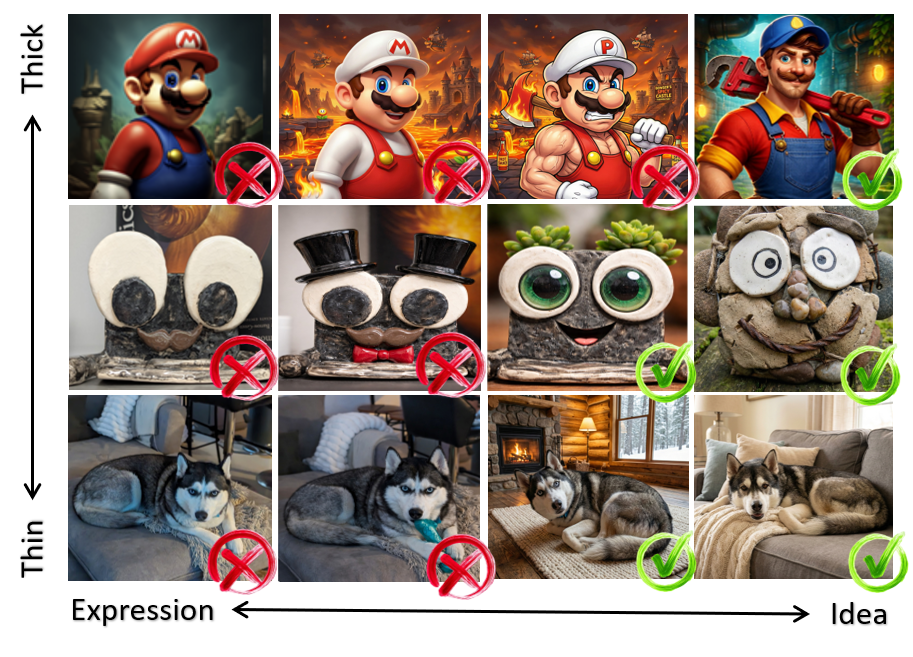}
    \vspace{-0.3cm}
    \caption{Illustration of the spectrum of copyright protection for generated images. Examples are arranged vertically from thin protection (bottom) to thick protection (top), and horizontally from expression (literal copies of the original image) to abstract idea. Except for the leftmost column, all variations were generated using Google’s Nano Banana 3 Pro model.}
    \label{fig:thin_v_thick_expamples}
    \vspace{-0.5cm}
\end{figure}

\subsection{Thick Protection: Character and Narrative-Driven}

Thick protection applies to works dominated by significant creative expression, such as fictional characters, distinctive visual designs, or developed narrative worlds. For such works, infringement is not limited to exact copying but extends to outputs that replicate protected expression. 

\paragraph{Non-Verbatim attacks} Near-duplicate extraction remains relevant here, as literal copying of a passage or image from a thickly protected work is still infringing. However, restricting analysis to verbatim reproduction is not necessary. \citet{cooper2025extracting}, for example, use ROUGE-style overlap rather than exact string matching, reflecting the fact that substantial similarity in expressive content can arise even when wording varies slightly. This illustrates that near-duplicate metrics already begin to blur into thick-protection concerns.  
\citet{copybench2024} show that LLMs can reproduce protected narrative and character structure under paraphrase, generating continuations that preserve distinctive characters, relationships, and fictional worlds even when no original wording is reused. What is reproduced in these cases is not a specific passage, but the expressive identity of the work itself. Because such elements receive thick copyright protection, their reconstruction can constitute infringement even in the absence of verbatim copying.

\paragraph{Copyright breach without memorization} More directly, text-to-image models have been shown to reconstruct protected expressive identities without reproducing any particular training image. \citet{beasts2025} demonstrate that diffusion models generate recognizable copyrighted characters from indirect or generic prompts, preserving defining visual attributes across poses, styles, and scenes. The reproduced content is not a specific image, but the character identity itself. Figure~\ref{fig:thin_v_thick_expamples} (top row) shows a character-level example (Super Mario) drawn from the same work discussed above, where none of the shown images are (to our knowledge) present in the training data. While the abstract idea of a video-game plumber is unprotected, outputs that remain clearly recognizable as the same character can infringe even after altering visual markers, pose, style, or scene. 

This mechanism is analyzed by \citet{chiba2025tackling}, which frame character generation as an originality–genericity problem. They show that characters with distinctive visual features can be reliably reconstructed across many prompts, yielding outputs that are consistently recognizable as the same copyrighted character despite substantial variation. This constitutes a character-level reconstruction attack: protected expressive identity is reproduced under variation rather than through literal duplication.

\paragraph{Strong protection is not forever.} Thick protection, however, is not absolute. As expressive elements become generic, standardized, or functional, their scope of protection weakens. Once expressive elements function as shared infrastructure rather than as creative choices, they may fall outside the core of copyright protection \citep{googleoracle2021}. Some work such as by  \citet{yarkoni2025} has demonstrated memorization of stock  visual elements such as  a generic mannequin used to "hang" t-shirts, or standardized e-commerce images of computer screens, which may have arguebly become functional and generic in this context, and, thereby, likely beyond the scope of copyright protection. Similarly, in \citet{Getty_v_StabilityAI_2025,USCO_Circular33_CFR2021}, Getty claimed that Stable Diffusion tends to memorize the company's Watermark, another functional feature which is not protected by copyright. 

Crucially, popularity alone does not eliminate protection. Distinctive characters and narrative worlds can remain strongly protected even when widely known, because their defining features remain creative and specific \citep{beasts2025,copybench2024}. Copyright decays only when expression becomes generic or functional, such as in idioms \cite{haviv2023understanding}, not merely when it becomes familiar.
\section{Call for Action}
We identify several key friction points where future research can contribute to a more principled formalization.

\subsection{Theory}

We stated our position that existing reconstruction attacks are not appropriate to make claims about whether a model contains a ``copy" of the data. This does not imply that models can never contain copies of data. Indeed, there are clear counterexamples: for instance, the nearest-neighbor algorithm is a discriminative model whose parameters are the data itself. Our position is consistent with this view.

We believe a correct theoretical framing of what constitute a copyright attack can contribute to clarify this separation. The theoretical framework of copyright attacks should provide a neutral formalism, describing attacks that are similar to an oblivious PDF reader and don't allow reconstruction from fragmented pieces of information using prior knowledge.  With the correct foramlism, the necessary mitigation methods can further be developed such as, for example, \emph{obfuscation} techniques in the context of cryptography and differential privacy in the context of privacy.
Providing the correct theoretical formalism can be very challenging, but it can be an interesting theoretical challenge, to provide a formal framework and assumptions underlying attacks that can correctly be interpreted and relevant in this context.

\subsection{Experimental Framing}

We reviewed several papers that expose threats to copyright material. Our position is that such result should be better framed. Exposing that copyrighted material has been memorized is a first step, but claim of relevance to copyright should be backed by correct reporting, and should be scrutinized according to the correct standards. In particular:
\begin{itemize}[leftmargin=9pt, nosep]
    \item \textbf{Copy vs. Copyrighted:} Whether the extracted or reconstructed outputs are substantially similar to protected works in a manner that would plausibly constitute copying in a copyright-relevant sense.
    \item \textbf{Information vs. Copying:} What prior knowledge was used during the attack. For example, access to the training data, to the copyrighted material, a trained model etc. 
    \item \textbf{Thin vs. Thick:} whether the success of the attack relied on the \emph{genericity} or widespread nature of the data, rather than on model-specific memorization;
\end{itemize}

For a generated output, a copyright-relevant evaluation should address the following questions: (i) Is the output substantially similar to protected expressions? (ii) Is the similarity literal or non-literal? (iii) Does similarity persist across diverse prompts? (iv) Would similar outputs arise without access to the original work?

Naturally, not every attack must adhere to all these standards to be relevant. However, as is standard in science, weaknesses should be reported and judged accordingly. Better framing can help guide \emph{next-step} research and inform how specific techniques should be further developed. 

\subsection{Technological Solutions}

The distinction between what constitutes copying and what does not hinges on subtle, context-dependent differences and often ambiguous boundaries. A possible mitigation strategy could be to harness generative AI to provide more insight into some of these questions. Generative AI models can be a tool, themselves to evaluate originality, genericity, and the expressive elements in a work -- properties which are not always clearly defined but require certain modeling of the distribution of the data. For example, something is generic if it frequently appears in several body of works. 

Several works have taken this pursuit, \citet{haviv2024not, hacohen2024, chiba2025tackling} propose leveraging generative AI models themselves to assess originality and generalization. The underlying hypothesis is that models trained on massive corpora provide a reasonable proxy of real-life distribution. Under this view, memorization is treated not as a failure mode but as a diagnostic tool, and treats collapse to memorization as signal for generic, non-expressive elements. 

This perspective offers two advantages: first, it naturally reflects variation in copyright scope across works, distinguishing thinly protected generic material from thickly protected expressive content; second, it shifts the focus away from memorization per se toward whether generated outputs collapse onto protected expressive identity.

\subsection{Benchmarking}

One class of mitigation approaches centers on \emph{benchmark-based evaluation} of model outputs. In this line of work, benchmarks are used not as definitions of infringement, but as diagnostic instruments that characterize output behaviors which may plausibly implicate copyright concerns.

Existing benchmarks vary considerably in the behaviors they target. They may focuse on \emph{literal or near-verbatim reproduction} of copyrighted text \cite{copybench2024,karamolegkou2023copyright}, or operate at the image level to evaluate when prompts omit explicit identifiers if outputs differ substantially from any single reference instance \cite{beasts2025}. This setting more closely reflects contexts of thick protection. Finally, a complementary strand of work evaluates \emph{model behavior under copyright-sensitive prompts}, measuring refusal rates, transformations, or other compliance-oriented responses rather than reproduction itself \cite{wei2024cotaval,mueller2024compliance}.

Crucially, benchmark outcomes should be interpreted as signals for mitigation rather than determinations of infringement. Copyright liability depends on multiple factors including substantial similarity of protected expression, access, and the availability of defenses such as fair use that are not fully captured by any single benchmark or metric.

\section{Alternative Views}

\paragraph{Generative AI are a copy of the data:}We caution against treating memorization attacks as evidence of a copy in the model. By contrast, \citet{cooper2025files} argue that “when a model has memorized training data, the model is a ‘copy’ of that training data \emph{in the sense used by copyright law}.” We instead assess copyright risk at the output level, based on how memorization manifests.

\paragraph{Memorization is often treated as an indicator of copyright risk.}
A widespread assumption in the ML literature is that a model’s ability to reproduce training data, particularly rare or copyrighted material, signals copyright risk. 
\cite{cooper2025extracting, yarkoni2025,somepalli2022diffusion,nasr2023scalable,karamolegkou2023copyright}. Our disagreement is with the legal interpretation. As discussed earlier, memorization frequently targets generic, functional, or weakly protected material, leading to systematic overestimation of infringement risk. At the same time, memorization-based tests may fail to capture non-literal reproduction of highly protected expression \cite{copybench2024,beasts2025,chiba2025tackling}.

\paragraph{Copyright concerns are sometimes framed primarily around training data.}
Another influential perspective emphasizes the legality of training on copyrighted data, independent of the content of generated outputs. From this viewpoint, memorization is concerning because it evidences internal retention of copyrighted works without authorization \cite{elkin2024,bracha2025,nyt2024}. This framing is common in policy debates and litigation, and reflects legitimate concerns about consent, compensation, and large-scale automated copying.

Our position is that these concerns, while important, are analytically distinct from copyright infringement. Copyright law regulates unauthorized reproduction and distribution of protected expression, not internal representations or latent influence \cite{lemleycasey2021,menell1989,googleoracle2021}. Questions of data governance, remuneration, or market substitution may warrant separate regulatory or contractual solutions, but conflating them with infringement doctrine risks obscuring the output-level legal standards courts apply.

\paragraph{Scope and Jurisdiction Disclaimer} While not an alternative view, we acknowledge that the legal analysis in this work primarily focuses on U.S. copyright law and does not claim direct applicability to other jurisdictions.

\nocite{langley00}

\paragraph{Acknowledgments}
This work was supported by the European Union (ERC, FoG 101116258). Views and opinions expressed are however those of the author(s) only and do not necessarily reflect those of the European Union or the European Research Council Executive Agency. Neither the European Union nor the granting authority can be held responsible for them.

Shay Moran is a Robert J.\ Shillman Fellow; he acknowledges support by ISF grant 1225/20, by BSF grant 2018385, by Israel PBC-VATAT, by the Technion Center for Machine Learning and Intelligent Systems (MLIS), and by the the European Union (ERC, GENERALIZATION, 101039692). Views and opinions expressed are however those of the author(s) only and do not necessarily reflect those of the European Union or the European Research Council Executive Agency. Neither the European Union nor the granting authority can be held responsible for them.
\bibliography{example_paper}

@inproceedings{langley00,
 author    = {P. Langley},
 title     = {Crafting Papers on Machine Learning},
 year      = {2000},
 pages     = {1207--1216},
 editor    = {Pat Langley},
 booktitle     = {Proceedings of the 17th International Conference
              on Machine Learning (ICML 2000)},
 address   = {Stanford, CA},
 publisher = {Morgan Kaufmann}
}

@misc{CopyrightWinter,
  author    = {Matthew Sag},
  title     = {Copyright Winter is Coming (to Wikipedia?)},
  year      = {2025},
  url       = {https://authorsalliance.substack.com/p/copyright-winter-is-coming-to-wikipedia},
}

@article{cooper2025extracting,
  title={Extracting memorized pieces of (copyrighted) books from open-weight language models},
  author={Cooper, A Feder and Gokaslan, Aaron and Ahmed, Ahmed and Cyphert, Amy B and De Sa, Christopher and Lemley, Mark A and Ho, Daniel E and Liang, Percy},
  journal={arXiv preprint arXiv:2505.12546},
  year={2025}
}

@article{litman1990,
  author = {Jessica Litman},
  title = {The Public Domain},
  journal = {Emory Law Journal},
  volume = {39},
  pages = {965--1023},
  year = {1990}
}

@article{menell1989,
  author  = {Menell, Peter S.},
  title   = {An Analysis of the Scope of Copyright Protection for Computer Programs},
  journal = {Stanford Law Review},
  volume  = {41},
  number  = {5},
  pages   = {1045--1103},
  year    = {1989}
}

@article{hacohen2024,
  author = {Uri Hacohen and Niva Elkin-Koren},
  title = {Genericity and the Scope of Copyright Protection},
  journal = {Journal of Law and Technology},
  year = {2024}
}

@inproceedings{feldman2020does,
  title={Does learning require memorization? a short tale about a long tail},
  author={Feldman, Vitaly},
  booktitle={Proceedings of the 52nd annual ACM SIGACT symposium on theory of computing},
  pages={954--959},
  year={2020}
}

@article{carlini2019secret,
  title={The Secret Sharer: Evaluating and Testing Unintended Memorization in Neural Networks},
  author={Carlini, Nicholas and Liu, Chang and Erlingsson, {\'U}lfar and Kos, Jernej and Song, Dawn},
  journal={USENIX Security},
  year={2019}
}

@article{carlini2021extract,
  title={Extracting Training Data from Large Language Models},
  author={Carlini, Nicholas and Tramer, Florian and Wallace, Eric and Jagielski, Matthew and Herbert-Voss, Ariel and Lee, Katherine and Roberts, Adam and Brown, Tom and Song, Dawn and Erlingsson, {\'U}lfar},
  journal={USENIX Security},
  year={2021}
}

@article{chang2023speak,
  title={Speak, Memory: An Archaeology of Books Known to ChatGPT/GPT-4},
  author={Chang, Kent K. and Bamman, David and others},
  journal={arXiv preprint arXiv:2305.00118},
  year={2023}
}

@article{zhao2024pip,
  title={Measuring Copyright Risks of Large Language Models via Partial Information Probing},
  author={Zhao, Wenyi and others},
  journal={arXiv preprint arXiv:2409.13831},
  year={2024}
}

@article{nasr2023scalable,
  title={Scalable Extraction of Training Data from (Production) Language Models},
  author={Nasr, Milad and Carlini, Nicholas and others},
  journal={IEEE S\&P},
  year={2023}
}

@article{karamolegkou2023copyright,
  title={Copyright Violations and Large Language Models},
  author={Karamolegkou, Antonia and others},
  journal={EMNLP},
  year={2023}
}

@article{somepalli2022diffusion,
  title={Diffusion Art or Digital Forgery? Investigating Data Replication in Diffusion Models},
  author={Somepalli, Gowthami and Singla, Vasu and Goldblum, Micah and Geiping, Jonas and Goldstein, Tom},
  journal={CVPR},
  year={2023}
}

@article{kasiviswanathan2011can,
  title={What can we learn privately?},
  author={Kasiviswanathan, Shiva Prasad and Lee, Homin K and Nissim, Kobbi and Raskhodnikova, Sofya and Smith, Adam},
  journal={SIAM Journal on Computing},
  volume={40},
  number={3},
  pages={793--826},
  year={2011},
  publisher={SIAM}
}

@article{floyd1995sample,
  title={Sample compression, learnability, and the Vapnik-Chervonenkis dimension},
  author={Floyd, Sally and Warmuth, Manfred},
  journal={Machine learning},
  volume={21},
  number={3},
  pages={269--304},
  year={1995},
  publisher={Springer}
}

@inproceedings{bassily2018learners,
  title={Learners that use little information},
  author={Bassily, Raef and Moran, Shay and Nachum, Ido and Shafer, Jonathan and Yehudayoff, Amir},
  booktitle={Algorithmic Learning Theory},
  pages={25--55},
  year={2018},
  organization={PMLR}
}

@inproceedings{mcallester1999pac,
  title={PAC-Bayesian model averaging},
  author={McAllester, David A},
  booktitle={Proceedings of the twelfth annual conference on Computational learning theory},
  pages={164--170},
  year={1999}
}

@article{feldman2025trade,
  title={Trade-offs in Data Memorization via Strong Data Processing Inequalities},
  author={Feldman, Vitaly and Kornowski, Guy and Lyu, Xin},
  journal={arXiv preprint arXiv:2506.01855},
  year={2025}
}

@misc{carlini2025privacy,
 title={privacy-copyright-and-generative-models},
  author= {Nicolas Carlini},
  url= {https://nicholas.carlini.com/writing/2025/privacy-copyright-and-generative-models.html},
 year={2025}
}

@article{carlini2023diffusion,
  title={Extracting Training Data from Diffusion Models},
  author={Carlini, Nicholas and Wallace, Eric and others},
  journal={USENIX Security},
  year={2023}
}

@article{copybench2024,
  title={CopyBench: Measuring Literal and Non-Literal Reproduction of Copyright-Protected Text in Language Model Generation},
  author={Chen, Zhen and others},
  journal={arXiv preprint arXiv:2407.07087},
  year={2024}
}

@inproceedings{beasts2025,
  title={Fantastic Copyrighted Beasts and How (Not) to Generate Them},
  author={He, Luxi and Huang, Yangsibo and Shi, Weijia and Xie, Tinghao and Liu, Haotian and Wang, Yue and Zettlemoyer, Luke and Zhang, Chiyuan and Chen, Danqi and Henderson, Peter},
  booktitle={The Thirteenth International Conference on Learning Representations},
  year={2024}
}

@inproceedings{haviv2023understanding,
  title={Understanding transformer memorization recall through idioms},
  author={Haviv, Adi and Cohen, Ido and Gidron, Jacob and Schuster, Roei and Goldberg, Yoav and Geva, Mor},
  booktitle={Proceedings of the 17th Conference of the European Chapter of the Association for Computational Linguistics},
  pages={248--264},
  year={2023}
}

@article{haviv2024not,
  title={Not every image is worth a thousand words: Quantifying originality in stable diffusion},
  author={Haviv, Adi and Sarfaty, Shahar and Hacohen, Uri and Elkin-Koren, Niva and Livni, Roi and Bermano, Amit H},
  journal={arXiv preprint arXiv:2408.08184},
  year={2024}
}

@article{chiba2025tackling,
  title={Tackling copyright issues in AI image generation through originality estimation and genericization},
  author={Chiba-Okabe, Hiroaki and Su, Weijie J},
  journal={Scientific Reports},
  volume={15},
  number={1},
  pages={10621},
  year={2025},
  publisher={Nature Publishing Group UK London}
}

@article{lemleycasey2021,
  title={Fair Learning},
  author={Lemley, Mark A. and Casey, Bryan},
  journal={Texas Law Review},
  year={2021}
}

@article{samuelson2024fairuse,
  author  = {Samuelson, Pamela},
  title   = {Fair Use Defenses in Disruptive Technology Cases},
  journal = {UCLA Law Review},
  volume  = {71},
  pages   = {1557--1559},
  year    = {2024},
  note    = {Available at SSRN: 4631726},
}

@article{wei2024cotaval,
  title={Evaluating Copyright Takedown Methods for Language Models},
  author={Wei, Boyi and others},
  journal={NeurIPS},
  year={2024}
}

@article{mueller2024compliance,
  title={LLMs and Memorization: On Quality and Specificity of Copyright Compliance},
  author={Mueller, Felix B. and others},
  journal={arXiv preprint arXiv:2405.18492},
  year={2024}
}

@inproceedings{elkin2024,
  author    = {Elkin-Koren, Niva and Hacohen, Uri and Livni, Roi and Moran, Shai},
  title     = {Can Copyright Be Reduced to Privacy?},
  booktitle = {Proceedings of the 5th Symposium on Foundations of Responsible Computing},
  year      = {2024},
}

@article{bracha2025,
  title={Generative AI’s Two Information Goods},
  author={Bracha, Oren},
  journal={Harvard Journal of Law \& Technology},
  year={2025}
}

@article{bracha2023work,
  title={The Work of Copyright in the Age of Machine Production},
  author={Bracha, Oren},
  journal={U of Texas Law, Legal Studies Research Paper},
  year={2023}
}

@article{nyt2024,
  title={Exploring Memorization and Copyright Violation in Frontier LLMs: A Study of the New York Times v. OpenAI Lawsuit},
  author={Anonymous},
  journal={arXiv preprint arXiv:2412.06370},
  year={2024}
}

@article{yarkoni2025,
  title={Low Resource Reconstruction Attacks Through Benign Prompts},
  author={Yarkoni, Sol and Sharif, Mahmood and Livni, Roi},
  journal={arXiv preprint arXiv:2507.07947},
  year={2025}
}

@inproceedings{attias2024information,
  title={Information complexity of stochastic convex optimization: Applications to generalization, memorization, and tracing},
  author={Attias, Idan and Dziugaite, Gintare Karolina and Haghifam, Mahdi and Livni, Roi and Roy, Daniel M},
  booktitle={Forty-first International Conference on Machine Learning},
  year={2024}
}

@article{beliy2019voxels,
  title={From voxels to pixels and back: Self-supervision in natural-image reconstruction from fMRI},
  author={Beliy, Roman and Gaziv, Guy and Hoogi, Assaf and Strappini, Francesca and Golan, Tal and Irani, Michal},
  journal={Advances in Neural Information Processing Systems},
  volume={32},
  year={2019}
}

@article{livni2023information,
  title={Information theoretic lower bounds for information theoretic upper bounds},
  author={Livni, Roi},
  journal={Advances in Neural Information Processing Systems},
  volume={36},
  pages={37716--37727},
  year= {2023}
}

@article{cooper2025files,
  title={The files are in the computer: on copyright, memorization, and generative AI},
  author={Cooper, A Feder and Grimmelmann, James},
  journal={Chi.-Kent L. Rev.},
  volume={100},
  pages={141},
  year={2025},
  publisher={HeinOnline}
}

@article{tishby2000information,
  title={The information bottleneck method},
  author={Tishby, Naftali and Pereira, Fernando C and Bialek, William},
  journal={arXiv preprint physics/0004057},
  year={2000}
}

@article{xu2017information,
  title={Information-theoretic analysis of generalization capability of learning algorithms},
  author={Xu, Aolin and Raginsky, Maxim},
  journal={Advances in neural information processing systems},
  volume={30},
  year={2017}
}

@inproceedings{brown2021memorization,
  title={When is memorization of irrelevant training data necessary for high-accuracy learning?},
  author={Brown, Gavin and Bun, Mark and Feldman, Vitaly and Smith, Adam and Talwar, Kunal},
  booktitle={Proceedings of the 53rd annual ACM SIGACT symposium on theory of computing},
  pages={123--132},
  year={2021}
}

@article{wen2018neural,
  title={Neural encoding and decoding with deep learning for dynamic natural vision},
  author={Wen, Haiguang and Shi, Junxing and Zhang, Yizhen and Lu, Kun-Han and Cao, Jiayue and Liu, Zhongming},
  journal={Cerebral cortex},
  volume={28},
  number={12},
  pages={4136--4160},
  year={2018},
  publisher={Oxford University Press}
}

@article{shen2019deep,
  title={Deep image reconstruction from human brain activity},
  author={Shen, Guohua and Horikawa, Tomoyasu and Majima, Kei and Kamitani, Yukiyasu},
  journal={PLoS computational biology},
  volume={15},
  number={1},
  pages={e1006633},
  year={2019},
  publisher={Public Library of Science San Francisco, CA USA}
}

@article{shen2019end,
  title={End-to-end deep image reconstruction from human brain activity},
  author={Shen, Guohua and Dwivedi, Kshitij and Majima, Kei and Horikawa, Tomoyasu and Kamitani, Yukiyasu},
  journal={Frontiers in computational neuroscience},
  volume={13},
  pages={21},
  year={2019},
  publisher={Frontiers Media SA}
}

@article{sag2023fairness,
  title={Fairness and fair use in generative AI},
  author={Sag, Matthew},
  journal={Fordham L. Rev.},
  volume={92},
  pages={1887},
  year={2023},
  publisher={HeinOnline}
}

@article{gillotte2019copyright,
  title={Copyright infringement in AI-generated artworks},
  author={Gillotte, Jessica L},
  journal={UC Davis L. Rev.},
  volume={53},
  pages={2655},
  year={2019},
  publisher={HeinOnline}
}

@article{lim2018ai,
  title={AI \& IP: innovation \& creativity in an age of accelerated change},
  author={Lim, Daryl},
  journal={Akron L. Rev.},
  volume={52},
  pages={813},
  year={2018},
  publisher={HeinOnline}
}

@article{grimmelmann2015copyright,
  title={Copyright for literate robots},
  author={Grimmelmann, James},
  journal={Iowa L. Rev.},
  volume={101},
  pages={657},
  year={2015},
  publisher={HeinOnline}
}

@article{Dornis_Lucchi_2025_GenAI_EU_Copyright,
  author  = {Dornis, Tim W. and Lucchi, Nicola},
  title   = {Generative AI and the Scope of EU Copyright Law: A Doctrinal Analysis in Light of C-250/25},
  journal = {International Review of Intellectual Property and Competition Law},
  volume  = {56},
  year    = {2025},
  note    = {Forthcoming, issue 10 (November 2025). Available at SSRN: 5391439},
}

@article{sag2020textmining,
  author  = {Sag, Matthew},
  title   = {The New Legal Landscape for Text Mining and Machine Learning},
  journal = {Journal of the Copyright Society of the USA},
  volume  = {66},
  pages   = {291--},
  year    = {2019},
  note    = {Available at SSRN: 3331606},
}

@book{Drassinower_2015_Whats_Wrong_with_Copying,
  author    = {Drassinower, Abraham},
  title     = {What’s Wrong with Copying?},
  publisher = {Harvard University Press},
  year      = {2015},
  note      = {p.~87},
}

@article{Craig_2025_AI_Copyright_Trap,
  author  = {Craig, Carys J.},
  title   = {The AI-Copyright Trap},
  journal = {Chicago-Kent Law Review},
  volume  = {100},
  pages   = {107--},
  year    = {2025},
  url     = {https://scholarship.kentlaw.iit.edu/cklawreview/vol100/iss1/8},
}

@incollection{Litman_2017_Fetishizing_Copies,
  author    = {Litman, Jessica},
  title     = {Fetishizing Copies},
  booktitle = {Copyright in an Age of Limitations and Exceptions},
  editor    = {Okediji, Ruth},
  publisher = {Cambridge University Press},
  year      = {2017},
  pages     = {107--131},
  url       = {https://scholarship.kentlaw.iit.edu/cklawreview/vol100/iss1/8},
}

@incollection{Strowel_2018_Reconstructing_Reproduction,
  author    = {Strowel, Alain},
  title     = {Reconstructing the Reproduction and Communication to the Public Rights: How to Align Copyright with its Fundamentals},
  booktitle = {Copyright Reconstructed},
  editor    = {Hugenholtz, P. Bernt},
  publisher = {Kluwer Law International},
  year      = {2018},
  pages     = {213--},
}

@misc{usc102b,
  author = {{17 U.S.C. § 102(b)}},
  title  = {Copyright Act, 17 U.S.C. § 102(b)},
  year   = {1976},
  note   = {Copyright Act of 1976},
}

@misc{usc107,
  author = {{17 U.S.C. § 107}},
  title  = {Copyright Act, 17 U.S.C. § 107},
  year   = {1976},
  note   = {Copyright Act of 1976 (Fair Use)},
}

@misc{usc106,
  author = {{17 U.S.C. § 106}},
  title  = {Copyright Act, 17 U.S.C. § 106},
  year   = {1976},
  note   = {Copyright Act of 1976 (Exclusive Rights in Copyrighted Works)},
}

@misc{usc101,
  author = {{17 U.S.C. § 101}},
  title  = {Copyright Act, 17 U.S.C. § 101},
  year   = {1976},
  note   = {Copyright Act of 1976 (Definitions)},
}

@misc{usco2021,
  author = {{U.S. Copyright Office}},
  title = {Compendium of U.S. Copyright Office Practices},
  year = {2021}
}

@misc{USCO_Circular33_CFR2021,
  author = {{U.S. Copyright Office}},
  title  = {Copyright Office Circular 33; 37 C.F.R. {\S}~202.1},
  year   = {2024},
  note   = {Copyright protection for works of the United States Government},
  url    = {https://www.copyright.gov/circs/circ33.pdf},
}

@misc{us_constitution,
  author = {{U.S. Const. art. I, § 8, cl. 8}},
  title  = {United States Constitution, Article I, Section 8, Clause 8},
  year   = {1787},
  note   = {Copyright Clause},
}

@misc{mazer1954,
  author       = {{Mazer v. Stein}},
  howpublished = {United States Supreme Court, 347 U.S. 201},
  year         = {1954},
  note         = {U.S. Supreme Court},
}

@misc{nichols1930,
  author       = {{Nichols v. Universal Pictures Corp.}},
  howpublished = {Federal Reporter, Second Series, Vol. 45, p. 119},
  year         = {1930},
  note         = {2d Cir.},
}

@misc{lotusborland1995,
  author       = {{Lotus Development Corp. v. Borland International, Inc.}},
  howpublished = {Federal Reporter, Third Series, Vol. 49, pp. 807--815},
  year         = {1995},
  note         = {1st Cir.},
}

@misc{lotusborland1996,
  author       = {{Lotus Development Corp. v. Borland International, Inc.}},
  howpublished = {United States Supreme Court, 516 U.S. 233},
  year         = {1996},
  note         = {U.S. Supreme Court},
}

@misc{cain1942,
  author       = {{Cain v. Universal Pictures Co.}},
  howpublished = {Federal Supplement, Vol. 47, p. 1013},
  year         = {1942},
  note         = {S.D. Cal.},
}

@misc{feist1991,
  author       = {{Feist Publications, Inc. v. Rural Telephone Service Co.}},
  howpublished = {United States Supreme Court, 499 U.S. 340},
  year         = {1991},
  note         = {U.S. Supreme Court},
}

@misc{campbell1994acuffrose,
  author       = {{Campbell v. Acuff-Rose Music, Inc.}},
  howpublished = {United States Supreme Court, 510 U.S. 569},
  year         = {1994},
  note         = {U.S. Supreme Court},
}

@misc{baker1879,
  author       = {{Baker v. Selden}},
  howpublished = {United States Supreme Court, 101 U.S. 99},
  year         = {1879},
  note         = {U.S. Supreme Court},
}

@misc{googleoracle2021,
  author        = {{Google LLC v. Oracle America, Inc.}},
  howpublished = {United States Supreme Court, 593 U.S.},
  year         = {2021},
  note         = {U.S. Supreme Court},
}

@misc{nash1990cbs,
  author = {{Nash v. CBS, Inc.}},
  title  = {Nash v. CBS, Inc.},
  year   = {1990},
  note   = {899 F.2d 1537 (7th Cir. 1990)},
}

@misc{mannion2005,
  author = {{Mannion v. Coors Brewing Co.}},
  title  = {Mannion v. Coors Brewing Co.},
  year   = {2005},
  note   = {377 F. Supp. 2d 444 (S.D.N.Y. 2005)},
}

@misc{applemicrosoft1994,
  author       = {{Apple Computer, Inc. v. Microsoft Corp.}},
  howpublished = {Federal Reporter, Third Series, Vol. 35, p. 1435},
  year         = {1994},
  note         = {9th Cir.},
}

@misc{Getty_v_StabilityAI_2025,
  author        = {{Getty Images (US) Inc. \& Ors v. Stability AI Ltd}},
  howpublished = {High Court of Justice of England and Wales, Chancery Division, [2025] EWHC 38 (Ch)},
  year         = {2025},
  note         = {Decision dated January 14, 2025},
}

@misc{Krofft_v_McDonalds_1977,
  author        = {Sid \& Marty Krofft Television Productions, Inc. v. McDonald’s Corp.},
  howpublished = {Federal Reporter, Second Series, Vol. 562, p. 1157},
  year         = {1977},
  note         = {9th Cir.},
}

@misc{GEMA_v_OpenAI_2024,
  author        = {{GEMA v. OpenAI}},
  howpublished = {Landgericht München I, I 42 O 14139/24},
  year         = {2024},
  note         = {Proceedings pending; action brought by the German collecting society GEMA against OpenAI},
}

@misc{CoStar_v_LoopNet_2004,
  author        = {{CoStar Group, Inc. v. LoopNet, Inc.}},
  howpublished = {Federal Reporter, Third Series, Vol. 373, p. 544},
  year         = {2004},
  note         = {4th Cir.},
}

@misc{Kadrey_v_Meta_2025,
  author        = {{Kadrey v. Meta Platforms, Inc.}},
  howpublished = {U.S. District Court for the Northern District of California, No. 3:23-cv-03417},
  year         = {2025},
  note         = {Order denying plaintiffs' motion for partial summary judgment and granting defendant's cross-motion for partial summary judgment; signed by Judge Vince Chhabria on June 25, 2025},
}

@misc{RTC_v_Netcom_1995,
  author        = {{Religious Technology Center v. Netcom On-Line Communication Services, Inc.}},
  howpublished = {907 F. Supp. 1361 (N.D. Cal.)},
  year         = {1995},
  note         = {United States District Court for the Northern District of California},
}

@misc{Bartz_v_Anthropic2025,
  author        = {{Bartz v. Anthropic PBC}},
  howpublished = {U.S. District Court for the Northern District of California, No. 3:24-cv-05417},
  year         = {2025},
  note         = {Filed Aug. 19, 2024; partial summary judgment on fair use issued June 23, 2025; class action copyright lawsuit alleging unauthorized use of authors’ works to train AI models and issues of fair use and piracy.},
  url          = {https://dockets.justia.com/docket/california/candce/3:2024cv05417/434709},
}

@misc{dccomics2015towle,
  author       = {{DC Comics v. Towle}},
  howpublished = {Federal Reporter, Third Series, Vol. 802, p. 1012},
  year         = {2015},
  note         = {9th Cir.},
}

@misc{ComputerAssoc_v_Altai_1992,
  author        = {{Computer Assocs. Int’l, Inc. v. Altai, Inc.}},
  howpublished = {Federal Reporter, Second Series, Vol. 982, pp. 693--707},
  year         = {1992},
  note         = {2d Cir.},
}

@misc{EtsHokin_v_SkyySpirits_2003,
  author       = {{Ets-Hokin v. Skyy Spirits, Inc.}},
  howpublished = {United States Court of Appeals for the Ninth Circuit},
  year         = {2003},
  note         = {9th Cir.},
}

@misc{UMG_v_MP3com_2000,
  author       = {{UMG Recordings, Inc. v. MP3.com, Inc.}},
  howpublished = {Federal Supplement, Second Series, Vol. 92, pp. 349--350},
  year         = {2000},
  note         = {S.D.N.Y.},
}
\bibliographystyle{icml2026}

\newpage
\appendix
\onecolumn
\section{Extended Copyright Background}
\label{apx:copyright_background_ext}

\cref{sec:copyright_background} provides a high-level copyright background. This appendix supplements that discussion with additional detail on infringement requirements and copyright scope.

\subsection{Which Copying Constitutes Copyright Infringement: Additional Details}
\label{apx:whichcopyiscopyright}

Building on ~\cref{subsec:whichcopyingconstitutes}, this subsection provides additional detail on the legal criteria that determine when copying is considered copyright infringement. Specifically, we elaborate on \textit{copying} as defined by law:

 The use involves reproduction of the copyrighted work “in \textit{copies}” (which are defined as: "material objects ... in which a work is fixed by any method now known or later developed, and from which the work can be perceived, reproduced, or otherwise communicated, either directly or with the aid of a machine or device" \cite{usc101}).  
    
A work is considered "fixed" when it's "embodiment in a copy ... is sufficiently permanent or stable to permit it to be perceived, reproduced, or otherwise communicated for a period of more than transitory duration" \cite{usc101}. 
    
Infringing copy requires \textit{access} to the copyrighted work and \textit{substantial similarity} to the protected elements within that work \cite{Krofft_v_McDonalds_1977}. 
  
\subsection{Copyright Scope and Infringement: Additional Details}
\label{apx:copyright_Infringement_background}
Several key principles govern the scope of copyright protection: 

\begin{enumerate}[leftmargin=10pt, nosep]
    \item \textbf{Facts and raw data are not protected} \cite{feist1991} For example, historical dates, scientific measurements, or biographical details. This principle was illustrated in the case of \citet{nash1990cbs} where an author claimed that a television show copied his historical research about John Dillinger. The court rejected the claim, holding that while Nash could own his expressive telling of history, the underlying facts and historical theories themselves remained free for others to reuse. 

    \item \textbf{Abstract ideas are excluded as distinct from their concrete expression} \cite{baker1879} For example the general idea of a detective story, a hero’s journey narrative arc, or an artistic style (Surrealism, Cubism) are unprotected \cite{nichols1930}. 

    \item \textbf{Functional elements and generic themes are not protected.} Such as procedures, systems, and methods of operation, are  excluded from copyright protection. The also includes standardized programming interfaces or formats required for interoperability. \cite{lotusborland1995,lotusborland1996} Similarly, generic cultural themes and narratives that are standard, indispensable, or customary to their genre or setting (scènes à faire),  are also unprotected \cite{cain1942}. Functionality and genericity are dynamic features that may change over time, as explained in \cref{it:scope}.  

    \item \textbf{Copyright law generally does not protect very small fragments of creative works.} As U.S.\ Copyright Office guidelines explain, ``words and short phrases, such as names, titles, and slogans,'' are uncopyrightable \cite{usco2021}. 

\item \label{it:scope} \textbf{The scope  of protection can vary} 
Even when copyright protection applies, its scope may vary substantially across different works. Creative works embody protected and unprotected elements, and the degree of protection afforded to the protected elements ranges from thin to thick \cite{feist1991, applemicrosoft1994}. The larger the ratio of protected elements compared to the non-protected elements in the copyrighted work the "thicker" the scope of legal protection over that work will be, and vice versa, the smaller the ratio of elements compared to the non-protected elements in the copyrighted work  the "thinner" the scope of legal protection over that work. These distinction apply across different domains such as textual data or visual data. 
\begin{itemize}[leftmargin=9pt,nosep]
    \item 
\textbf{Thin Protection} applies when expressive choices are heavily constrained by facts, functionality, or convention. In such cases, the scope of protection is thin, and only a near-identical copying of all the expressive details in the copyrighted work will constitute copyright infringement \cite{applemicrosoft1994}. Minor changes and variations might render the allegedly infringing use non-infringing, namely, beyond the scope of copyright protection. 

For example, in the context of language data, a news article that primarily conveys factual information receives thin copyright protection. Copying the underlying facts or paraphrasing them in different words does not infringe, whereas reproducing a distinctive paragraph verbatim may \cite{nash1990cbs}. In the context of images, a product photograph used for documentation or e-commerce typically receives thin protection based  on  original compositional choices, so only near-duplicate images would be considered infringement \cite{mannion2005}.

\item \textbf{Thick Protection} applies when a work reflects a high degree of original expressive choices, in which case infringement may extend to non-literal reproductions that capture its distinctive expressive structure. Fictional novels, for example, receive thick copyright protection. Even if the text is rephrased, reproducing distinctive characters, relationships, or narrative structure can constitute infringement. Thus, while a rephrased  synopsis of a news article is unlikely to infringe copyright in the original article, such a rephrased synopsis of an original novel  might \cite{CopyrightWinter}. The same principle  also applies to images: distinctive and sufficiently delineated fictional characters receive thick copyright protection that extends across various stories and media. Such characters enjoy thicker protection because they encapsulate many expressive choices concerning their distinctive appearance, narrative and identifiable characteristic traits \cite{dccomics2015towle}
\end{itemize}

\item \textbf{The scope of protection changes over time}
As expressive elements become widely adopted, standardized, or socially entrenched, they may become generic or acquire a functional character, weakening the scope of protection \cite{hacohen2024,menell1989}. A prominent example is the Supreme Court's decision in \emph{Google LLC v.\ Oracle America, Inc.}, which held that certain Java interface declarations, once widely adopted and functionally necessary for interoperability, were not protected against reuse \cite{googleoracle2021}. Analogously, visual templates that become standard in commercial imagery or interface design may acquire a functional role that narrows or eliminates protection. 

Crucially, not all expressive elements lose protection through familiarity. Some expressive constructs most notably well-developed fictional characters, may remain protected even when widely recognized. Copyright protection for such characters often extends beyond exact textual or visual depictions to encompass distinctive combinations of traits, attributes, and narrative roles.

\item \textbf{Even protected works may be used freely depending on context.} Unauthorized uses of protected copyright material may be excused from liability if they fall within the recognized exceptions or limitations to copyright protection, most notably "fair use" \cite{usc107}.  Fair use permits uses of copyrighted materials in contexts that are considered socially desirable, such as criticism, research, or parody \cite{campbell1994acuffrose}.

\item \label{apx: fair_use} \textbf{Fair use} privilege under U.S. copyright law is an open-ended, flexible legal standard, which empowers courts to carve out an exception for an otherwise infringing use after weighing a set of statutory factors (17 U.S.C. § 107):  (1) the purpose and character of the use, including whether such use is of a commercial nature or is for nonprofit educational purposes; (2) the nature of the copyrighted work; (3) the amount and substantiality of the portion used in relation to the copyrighted work as a whole; and
(4) the effect of the use upon the potential market for or value of the copyrighted work.
Importantly, a finding of fair use does not require satisfying all four factors, nor do the factors function as a mathematical formula where, for example, three factors in favor automatically determine the outcome. Instead, all four factors must be considered and weighed together, in light of the broader goals of copyright \cite{campbell1994acuffrose}. 

Reflecting this flexibility, §107 provides a non-exhaustive list of examples of uses that may qualify as fair use—such as “criticism, comment, news reporting, teaching (including multiple copies for classroom use), scholarship, [and] research.” Courts retain significant discretion to interpret and apply the fair use doctrine to evolving contexts, including new technologies.

\end{enumerate}




\end{document}